\documentclass[12pt]{article}
\usepackage{amsmath}
\usepackage{amssymb}
\usepackage{graphics}
\usepackage{epsf,graphicx,rotating,color}
\usepackage{bm,bbm}
\usepackage{amsfonts}

\def\r{\bf r}
\def\R{\bf R}

\begin{document}
\title{{\bf Two bodies gravitational system with variable mass and  damping-anti damping effect due to star wind}}
\author{ G.V.  L\'opez\footnote{gulopez@udgserv.cencar.udg.mx} and E. M. Ju\'arez\\\\Departamento de F\'{\i}sica de la Universidad de Guadalajara,\\
Blvd. Marcelino Garc\'{\i}a Barrag\'an 1421, esq. Calzada Ol\'{\i}mpica,\\
44430 Guadalajara, Jalisco, M\'exico03.67.Lx,\  03.67.Hr,\  03.67.-a,\  03.65w}
\date{July, 2010}
\maketitle

\begin{abstract}
We study two-bodies gravitational problem  where the mass of one of the bodies varies and suffers a damping-anti damping
effect due to star wind during its motion. A constant of motion, a Lagrangian and a Hamiltonian are given for the radial motion of the
system, and the period of the body is studied using the constant of motion of the system.  An application to the comet motion is given, using the comet Halley as an example.
\end{abstract}
\newpage
\section{\bf Introduction}
\noindent
There is not doubt that mass variable systems have been relevant since the foundation of the classical mechanics and
modern physics too [0] which have been known as  Gylden-Meshcherskii problems [1]. Among these type of systems one could mention: the motion of rockets [2], the kinetic theory
of dusty plasma [3], propagation of electromagnetic waves in a dispersive-nonlinear media [4], neutrinos mass 
oscillations [5], black holes formation [6], and comets interacting with solar wind [7]. This last system belong to the 
so called "gravitational two-bodies problem" which is one of the most studied and well known system in classical
mechanics [8]. In this type of system, one assumes normally that the masses of the bodies are fixed and unchanged
during the dynamical motion. However,when one is dealing with comets, beside to consider its mass variation due to the
interaction with the solar wind, one would like to have an estimation of the the effect of the solar wind pressure on the
comet motion. This pressure may produces a dissipative-antidissipative effect on its motion. The dissipation effect must
be felt by the comet when this one is approaching to the sun (or star), and the antidissipation effect must be felt by
the comet when this one is moving away from the sun. \\\\
\noindent
In previous paper [14]. a study was made of the two-bodies gravitational problem with mass variation in one of them, where we were interested in the difference of the  trajectories in the spaces ($x,v$) and ($x,p$).  In this paper, we study the two-bodies gravitational problem taking into consideration the mass variation of one of
them  and its damping-anti damping effect due to the solar wind. The mass of the other body is assumed big and fixed , and the
reference system of motion  is chosen just in this body. In addition, we will assume that the mass lost is expelled from the body radially to its motion. Doing this, the three-dimensional two-bodies problem is
reduced to a one-dimensional problem. Then, a constant of motion, the Lagrangian, and the Hamiltonian are deduced for  this one-dimensional problem, where a radial dissipative-antidissipative force proportional to the velocity square is
chosen. A model for the mass variation is given, and the damping-anti damping effect is studied on the period of the trajectories, the trajectories themselves, and the aphelion distance of a comet.   We use the parameters associated to comet Halley to illustrate the application of our results.
\newpage\noindent
\section{\bf  Equations of Motion.}
\noindent
Newton's equations of motion for two bodies interacting gravitationally, seen from arbitrary
inertial reference system, and with radial dissipative-antidissipative force acting in one of them are given by
\begin{subequations}
\begin{equation}
{d\over dt}\left(m_1{d{\r_1}\over dt}\right)=-{Gm_1m_2\over
|{\r_1}-{\r_2}|^3}(\r_1-\r_2)
\end{equation}
 and
 \begin{equation}
{d\over dt}\left(m_2{d{\r_2}\over dt}\right)=-{Gm_1m_2\over |{\r_2}-{\r_1}|^3}(\r_2-\r_1)
-{\gamma\over|\r_1-\r_2|}\left[{d|\r_1-\r_2|\over dt}\right]^2(\r_2-r_1)\ ,
\end{equation}
\end{subequations}
where $m_1$ and $m_2$ are the masses of the two bodies, ${\r_1}=(x_1,y_1,z_1)$ and \hfil\break
${\r_2}=(x_2,y_2,z_2)$ are their vectors positions from the reference system, $G$ is
the gravitational constant ($G=6.67\times 10^{-11}m^3/Kg~s^2$), $\gamma$ is the nonnegative constant parameter of the
dissipative-antidissipative force, and 
\begin{equation*} 
{|\r_1-\r_2|=|\r_2-\r_1|}=\sqrt{(x_2-x_1)^2+(y_2-y_1)^2+(z_2-z_1)^2}
\end{equation*}
is the Euclidean distance between the two bodies. Note that if $\gamma> 0$ and \\ $d|\r_1-\r_2|/dt>0$ one has dissipation since the force acts against the motion of the body, and for $d|\r_1-\r_2|/dt<0$ one has anti-dissipation since the force pushes the body. If $\gamma<0$ this scheme  is reversed and  corresponds to our actual situation with the comet mass lost. \\\\
\noindent
It will be assumed the mass $m_1$ of the first body is constant and that the mass $m_2$ of the second body varies. Now, It is clear that the usual
relative, $\r$, and center of mass, $\R$, coordinates defined as
${\r}={\r_2}-{\r_1}$ and ${\R}=(m_1{\r_1}+m_2{\r_2})/( m_1+m_2)$ are not so good to describe the dynamics of this system.
However, one can consider the case for $m_1\gg m_2$ (which is the case star-comet), and consider
to put our reference system just on the first body ($\r_1=\vec 0$). In this case, Eq. (1a) and Eq. (1b) are reduced to the equation
\begin{equation}
m_2{d^2{\r}\over dt^2}=-{Gm_1m_2\over r^3}~{\r}-\dot{m}_2\dot{\r}-{\gamma}\left[{dr\over dt}\right]^2\hat {\r}\ ,
\end{equation}
where one has made the definition ${\r}={\r_2}=(x,y,z)$,  $r$ is its magnitude, $r=\sqrt{x^2+y^2+z^2}$, and $\hat{\r}={\r}/r$ is the unitary radial vector.  Using
spherical coordinates ($r,\theta,\varphi$),
\begin{equation}
x=r\sin\theta\cos\varphi\ ,\ \ y=r\sin\theta\sin\varphi\ ,\ \ z=r\cos\theta\ ,
\end{equation}
one obtains the following coupled equations
\begin{equation}
m_2(\ddot{r}-r\dot{\theta}^2-r\dot{\varphi}^2\sin^2\theta)=-{Gm_1m_2\over r^2}-\dot{m}_2\dot{r}-\gamma\dot{r}^2\
,\end{equation}
\begin{equation}
m_2(2\dot{r}\dot{\theta}+r\ddot{\theta}-r\dot{\varphi}^2\sin\theta\cos\theta)=
-\dot{m}_2r\dot{\theta}\ ,\end{equation}
and
\begin{equation}
m_2(2\dot{r}\dot{\varphi}\sin\theta+r\ddot{\varphi}\sin\theta+2r\dot{\varphi}\dot{\theta}\cos\theta)
=-\dot{m}_2r\dot{\varphi}\sin\theta\ .\end{equation}
Taking $\dot{\varphi}=0$  as solution of this last equation,  the resulting equations are
\begin{equation}
m_2(\ddot{r}-r\dot{\theta}^2)=-{Gm_1m_2\over r^2}-\dot{m}_2\dot{r}-\gamma\dot{r}^2\ ,
\end{equation}
and
\begin{equation}\label{8}
m_2(2\dot{r}\dot{\theta}+r\ddot{\theta})+\dot{m}_2r\dot{\theta}=0\ .
\end{equation}
From this last expression,   one  gets the following constant of motion (usual angular momentum of the system) 
\begin{equation}\label{9}
l_{\theta}=m_2r^2\dot{\theta}\ ,
\end{equation}
and with this constant of motion substituted in  Eq. 7, one obtains the following one-dimensional equation of motion for the radial part 
\begin{equation}\label{10}
{d^2r\over dt^2}=-{Gm_1\over r^2}-{\dot{m}_2\over m_2}\left({dr\over dt}\right)
-{\gamma\over m_2}\dot{r}^2+{l_{\theta}^2\over m_2^2r^3}\ .
\end{equation}
Now, let us assume that $m_2$ is a function of the distance between the first and the second body, $m_2=m_2(r)$.
Therefore, it follows that 
\begin{equation}\label{11}
\dot{m}_2=m_2'\dot{r}\ ,
\end{equation}
where $m_2'$ is defined as $m_2'=dm_2/dr$. Thus, Eq. (10) is written as
\begin{equation}\label{12}
{d^2r\over dt^2}=-{Gm_1\over r^2}+{l_{\theta}^2\over m_2^2 r^3}
-{{m}_2'+\gamma\over m_2}\left({dr\over dt}\right)^2\ ,
\end{equation} 
which, in turns, can be  written as the following autonomous dynamical system 
\begin{equation}\label{13}
{dr\over dt}=v\ ;\quad\quad{dv\over dt}=-{Gm_1\over r^2}+{l_{\theta}^2\over m_2^2r^3}
-{{m}_2'+\gamma\over m_2}v^2\ .
\end{equation}
Note from this equation that $m_2'$ is always a non-positive function of $r$ since it represents the mass lost rate. On the other hand,  $\gamma$ is a negative parameter in our case. 
\section{\bf Constant of Motion, Lagrangian and Hamiltonian}
\noindent
A constant of motion for the dynamical system (13) is a function $K=K(r,v)$ which
satisfies the partial differential equation [9]
$$v{\partial K\over\partial r}+\left[{-Gm_1\over r^2}+{l_{\theta}^2\over m_2^2 r^3}
-{m_2'+\gamma\over m_2}v^2\right]{\partial K\over\partial v}=0\ .\eqno{(14)}$$
The general solution of this equation is given by [10]
$$K(x,v)=F(c(r,v))\ ,\eqno(15a)$$
where $F$ is an arbitrary function of the characteristic curve $c(r,v)$ which has the following expression
$$c(r,v)=m_2^2(r)e^{2\gamma\lambda(r)}v^2+\int\left({2Gm_1\over r^2}-{2l_{\theta}^2\over m_2^2r^3}\right)m_2^2(r) e^{2\gamma\lambda(r)}dr\ ,\eqno(15b)$$
and the function $\lambda(r)$ has been defined as
$$\lambda(r)=\int{dr\over m_2(r)}\ .\eqno(15c)$$
During a cycle of oscillation, the function $m_2(r)$ can be different for the comet approaching the sun and for the
comet moving away from the sun. Let us denote $m_{2+}(r)$ for the first case and $m_{2-}(r)$ for the second case.
Therefore, one has two cases to consider in Eqs. (15a), (15b) and (15c) which will denoted by ($\pm$). Now, if 
$m_{2\pm}^o$ denotes the mass at aphelium (+) or perielium (-) of the comet, $F(c)=c^{\pm}/2m_{2\pm}^o$ represents
the functionality in Eq. (15a) such that for $m_2$ constant and $\gamma$ equal zero, this constant of motion is the
usual gravitational energy. Thus, the constant of motion can be chosen as $K^{\pm}=c(r,v)/2m_{2\pm}^0$, that is, 
$$K^{\pm}={m_{2\pm}^2(r)\over 2m_{2\pm}^o}e^{2\gamma\lambda_{\pm}(r)}v^2+V_{eff}^{\pm}(r)\ ,\eqno(16a)$$
where the effective potential $V_{eff}$ has been defined as
$$
V_{eff}^{\pm}(r)={Gm_1\over m_{2\pm}^o}\int{m_{2\pm}^2(r)e^{2\gamma\lambda_{\pm}(r)}dr\over r^2}
-{l_{\theta}^2\over m_{2\pm}^o}       \int{e^{2\gamma\lambda_{\pm}(r)}dr\over r^3}
\eqno(16b)$$
This effective potential has an extreme at the point $r_*$ defined by the relation
$$r_*m_2^2(r_*)={l_{\theta}^2\over Gm_1}\eqno(17)$$
which is independent on the parameter $\gamma$ and depends on the behavior of $m_2(r)$. This extreme point is a minimum of the effective potential
since one has
$$\left({d^2V_{eff}^{\pm}\over dr^2}\right)_{r=r_*}>0\ .\eqno(18)$$
Using the known expression [11-13] for the Lagrangian in terms of the constant of motion,
$$L(r,v)=v\int{K(r,v)~dv\over v^2}\ ,\eqno(19)$$
the Lagrangian, generalized linear momentum and the Hamiltonian are given by
$$L^{\pm}={m_{2\pm}^2(r)\over 2m_{2\pm}^o}e^{2\gamma\lambda_{\pm}(r)}v^2-V_{eff}^{\pm}(r)\ ,\eqno(20)$$ 
$$p={m_{2\pm}^2(r)~v\over m_{2\pm}^o}e^{2\gamma\lambda_{\pm}(r)}\ ,\eqno(21)$$
and
$$H^{\pm}={m_{2\pm}^op^2\over 2m_{2\pm}^2(r)}e^{-2\gamma\lambda_{\pm}(r)}+V_{eff}^{\pm}(r)\ .\eqno(22)$$
The trajectories in the space ($x,v$) are determined by the constant of motion (16a). Given the initial condition
($r_o,v_o$), the constant of motion has the specific value
$$K^{\pm}_o={m_{2\pm}^2(r_o)\over 2m_{2\pm}^o}e^{2\gamma\lambda_{\pm}(r_o)}v^2_o+V_{eff}^{\pm}(r_o)\ ,\eqno(23)$$
and the trajectory in the space ($r,v$) is given by
$$v=\pm\sqrt{2m_{2\pm}^o\over m_{2\pm}^2(r)}~e^{-\gamma\lambda_{\pm}(r)}\biggl[K_o^{\pm}-V_{eff}^{\pm}(r)\biggr]^{1/2}
\ .\eqno(24)$$
Note that one needs to specify $\dot{\theta}_o$ also to determine Eq. (9). In addition, one normally wants to know
the trajectory in the real space, that is, the acknowledgment of $r=r(\theta)$. Since one has that
$v=dr/dt=(dr/d\theta)\dot\theta$ and Eqs. (9) and (24), it follows that
$$\theta(r)=\theta_o+{l_{\theta}^2\over\sqrt{2m_{2\pm}^o}}\int_{r_o}^r{m_{2\pm}(r)e^{\gamma\lambda_{\pm}(r)}dr
\over r^2\sqrt{K_o^{\pm}-V_{eff}^{\pm}(r)}}\ .\eqno(25)$$
The half-time period (going from aphelion to perihelion (+), or backward (-)) can be deduced from Eq. (24) as
$$T_{1/2}^{\pm}={1\over\sqrt{2m_{2\pm}^o}}\int_{r_1}^{r_2}{m_{2\pm}(r)e^{\gamma\lambda_{\pm}(r)}dr
\over \sqrt{K_o^{\pm}-V_{eff}^{\pm}(r)}}\ ,\eqno(26)$$
where $r_1$ and $r_2$ are the two return points  resulting from the solution of the following equation
$$V_{eff}^{\pm}(r_i)=K_o^{\pm}\quad i=1,2\ .\eqno(27)$$
On the other hand, the trajectory in the
space ($r,p$) is determine by the Hamiltonian (21), and given the same initial conditions, the initial $p_o$ and
$H_o^{\pm}$ are obtained from Eqs. (21) and  (22). Thus, this trajectory is given by
$$p=\pm\sqrt{2m_{2\pm}^2(r)\over m_{2\pm}^o}~e^{\gamma\lambda_{\pm}(r)}\biggl[H_o^{\pm}-V_{eff}^{\pm}(r)\biggr]^{1/2}
\ .\eqno(28)$$
It is clear just by looking the expressions (24) and (28) that the trajectories in the spaces ($r,v$) and ($r,p$) must be different due to complicated relation (21) between $v$ and $p$ (see reference [14]).\\\\
\section{\bf  Mass-Variable Model and Results}
\noindent
As a possible application, consider that a comet looses material as a result of
the interaction with star wind in the following way (for one cycle of oscillation)
\begin{equation*}
m_{2\pm}(r)=\Biggl\{\begin{array}{l l} 
m_{2-}(r_{2(i-1)})\biggl(1-e^{-\alpha r}\biggr)& incoming (+)\  v<0\\ \\
m_{2+}(r_{2i-1})-b\biggl(1-e^{-\alpha (r-r_{2i-1})}\biggr)& outgoing (-)\  v>0
 \end{array}
\end{equation*}
$$\eqno(29)$$
where the parameters $b>0$ and $\alpha>0$ can be chosen to math  the mass loss rate in the incoming and outgoing cases. The
index "i" represent the ith-semi-cycle, being $r_{2(i-1)} $ and $r_{2i-1} $ the aphelion($r_a$) and perihelion($r_p$) points ($r_o$ is given by the initial conditions, and one has that $m_{2-}(r_o)=m_o$). For this case, the functions $\lambda_{+}(r)$ and
$\lambda_{-}(r)$ are given by
$$\lambda_{+}(r)={1\over\alpha m_a}~\ln{\biggl(e^{\alpha r}-1\biggr)}\ ,\eqno(30a)$$
and
$$
\lambda_{-}(r)={ -1\over \alpha(b-m_p)}
\Biggl[\alpha r+\ln{\left(m_p-b(1-e^{-\alpha(r-r_p)})\right)}\Biggr]\ .\eqno(30b)$$
where we have defined $m_a=m_2(r_a)$ and $m_p=m_2(r_p)$. Using the Taylor expansion, one gets
$$
e^{2\gamma\lambda_+(r)}=e^{2\gamma r/m_a}\biggl[1-{2\gamma\over\alpha m_a} e^{-\alpha r}+{1\over 2}{2\gamma\over\alpha m_a}\left({2\gamma\over\alpha m_a}-1\right)~e^{-2\alpha r}+\dots\biggr]\ ,
\eqno(31a)$$
and
\begin{eqnarray*}
& &e^{2\gamma\lambda_-(r)}={e^{-{2\gamma r\over (b-m_p)}}\over (m_p-b)^{{2\gamma \over\alpha(m_p-b)}}}
\biggl[1+{2\gamma \over\alpha(m_p-b)}~{e^{-\alpha(r-r_p)}\over m_p-b}\\
& & \quad \quad \quad+
{1\over 2}{2\gamma \over\alpha(m_p-b)}\left({2\gamma \over\alpha(m_p-b)}-1\right)~{e^{-2\alpha(r-r_p)}\over (m_p-b)^2}+\dots\biggr]
\end{eqnarray*}
$$\eqno(31b)$$
The effective potential for
the incoming comet can  be written as
$$V_{eff}^+(r)=
\left[-\frac{Gm_1m_a}{r}+{l_{\theta}^2\over 2 m_a}\frac{1}{r^2}\right]
e^{2\gamma r/m_a}+W_1(\gamma,\alpha,r)\ ,
\eqno(32)$$
and for the outgoing comet as
$$V_{eff}^-(r)=
\left[-\frac{Gm_1m_a}{r}+{l_{\theta}^2\over 2 m_a}\frac{1}{r^2}\right]
{e^{{2\gamma r\over (m_p-b)}}\over (m_p-b)^{{2\gamma \over\alpha(m_p-b)}}}
+W_2(\gamma,\alpha,r)\ ,
\eqno(33)$$
where $W_1$ and $W_2$ are given in the appendix. \\\\
\noindent
We will use  the data corresponding to the sun mass ($1.9891\times 10^{30}Kg$) and the  Halley comet [15-17],
$$m_c\approx 2.3\times 10^{14}Kg,\quad r_p\approx 0.6~au,\quad r_a\approx 35~au, \quad l_{\theta}\approx 10.83\times 10^{29}Kg\cdot m^2/s, \eqno(34)$$
with a mass lost of about $\quad \delta m\approx 2.8\times10^{11}Kg$ per cycle of oscillation. Although, the behavior of Halley comet seem to be chaotic [18], but we will neglect this fine detail here.  Now,  the parameters $\alpha$ and $"b"$ appearing on the mass lost model, Eq. (29), are determined by the chosen mass lost of the comet during the approaching to the sun and during the moving away from  the sun (we have assumed the same mass lost in each half of the cycle of oscillation of the comet around the sun). 
Using Eqs. (32) and (33), Eq. (24), the trajectories can be calculated in the spaces ($r,v$) . Fig. 1 shows these
trajectories using  $\delta m=2\times 10^{10}Kg$ (or $\delta m/m=0.0087\%$) for  $\gamma=0$ and (continuos line), and for $\gamma=-3~Kg/m$ (dashed line), starting both cases from the same aphelion distance. As one can see on the minimum,  dissipation causes to reduce a little bit the velocity of the comet , and the antidissipation increases the comet velocity, reaching a further away aphelion point.  Also, when only mass lost is considered ($\gamma=0$) the comet returns to aphelion point a little further away from the initial one
during the cycle of oscillation. Something related with this effect is the change of period as a function of mass lost ($\gamma=0$). This can be see on Fig. 2, where the period is calculated starting always from 
the same aphelion point ($r_a$). Note that with a mass lost of the order $2.8\times10^{11}Kg$ (Halley comet), which correspond to $\delta m/m=.12\%$,  the comet is well within 75 years period. The variation of the ratio of the change of aphelion distance as a function of mass lost ($\gamma=0$) is shown on Fig.3.  On Fig. 4, the mass lost rate is kept fixed to $\delta m/m=0.0087\%$, and the variation of the period of the comet is calculated as a function of the dissipative-antidissipative parameter $\gamma<0$ (using $|\gamma|$ for convenience). As one can see, antidissipation always wins to dissipation, bringing about the increasing of the period as a function of this parameter. The reason seems to be that the  antidissipation acts on the comet when this ones is lighter than when dissipation was acting (dissipation acts when the comet approaches to the sun, meanwhile antidissipation acts when the comet goes away from the sun). Since the period of Halley comets has not changed much during many turns, we can assume that the parameter $\gamma$ must  vary in the interval $(-0.01,0] Kg/m$.
 Finally, Fig. 5 shows the variation, during a cycle of oscillation, of the ratio of the new aphelion ($r_a'$) to old aphelion ($r_a$) as a function of the parameter $\gamma$.\\\\
\noindent

\begin{figure}[H]
 \begin{center}
 \includegraphics[width=8.5cm,height=6cm]{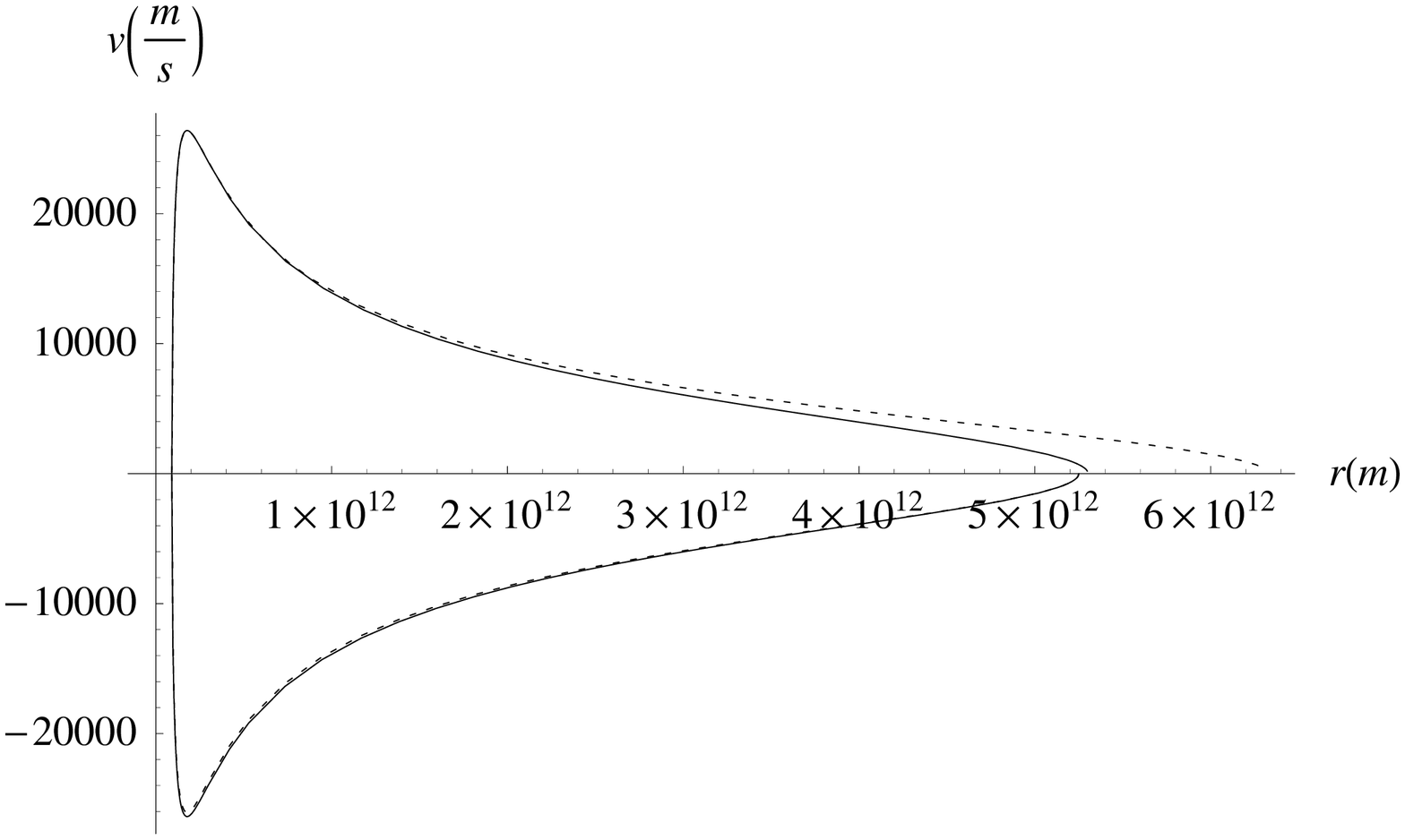}
 \end{center}
 \caption{ Trajectories in the ($r,v$) space with $\delta m/m=0.009$.}
\end{figure}

\begin{figure}[H]
 \begin{center}
 \includegraphics[width=10.5cm,height=8cm]{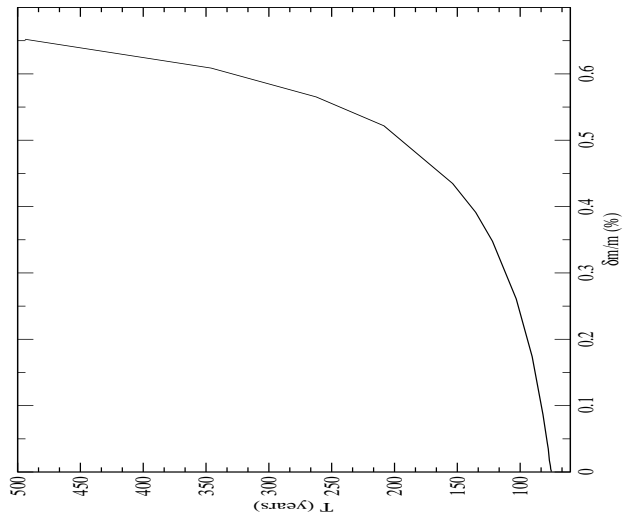}
 \end{center}
 \caption{ Period of the comet as a function of the mass lost ratio.}
\end{figure}

\begin{figure}[H]
 \begin{center}
 \includegraphics[width=8.5cm,height=6cm]{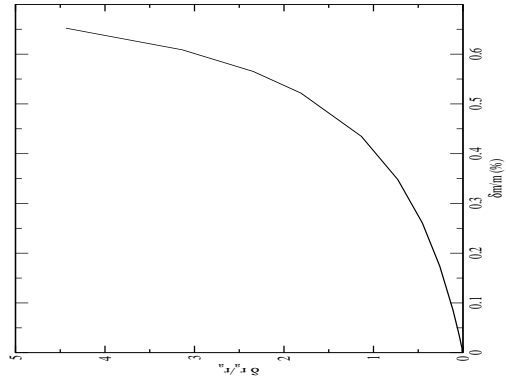}
 \end{center}
 \caption{ Ratio of aphelion distance change as a function of the mass lost rate.}
\end{figure}

\begin{figure}[H]
 \begin{center}
 \includegraphics[width=8.5cm,height=6cm]{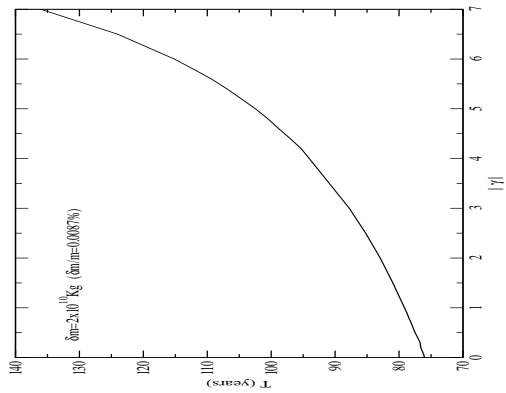}
 \end{center}
 \caption{Period of the comet as a function of the parameter $\gamma$.}
\end{figure}

\begin{figure}[H]
 \begin{center}
 \includegraphics[width=8.5cm,height=6cm]{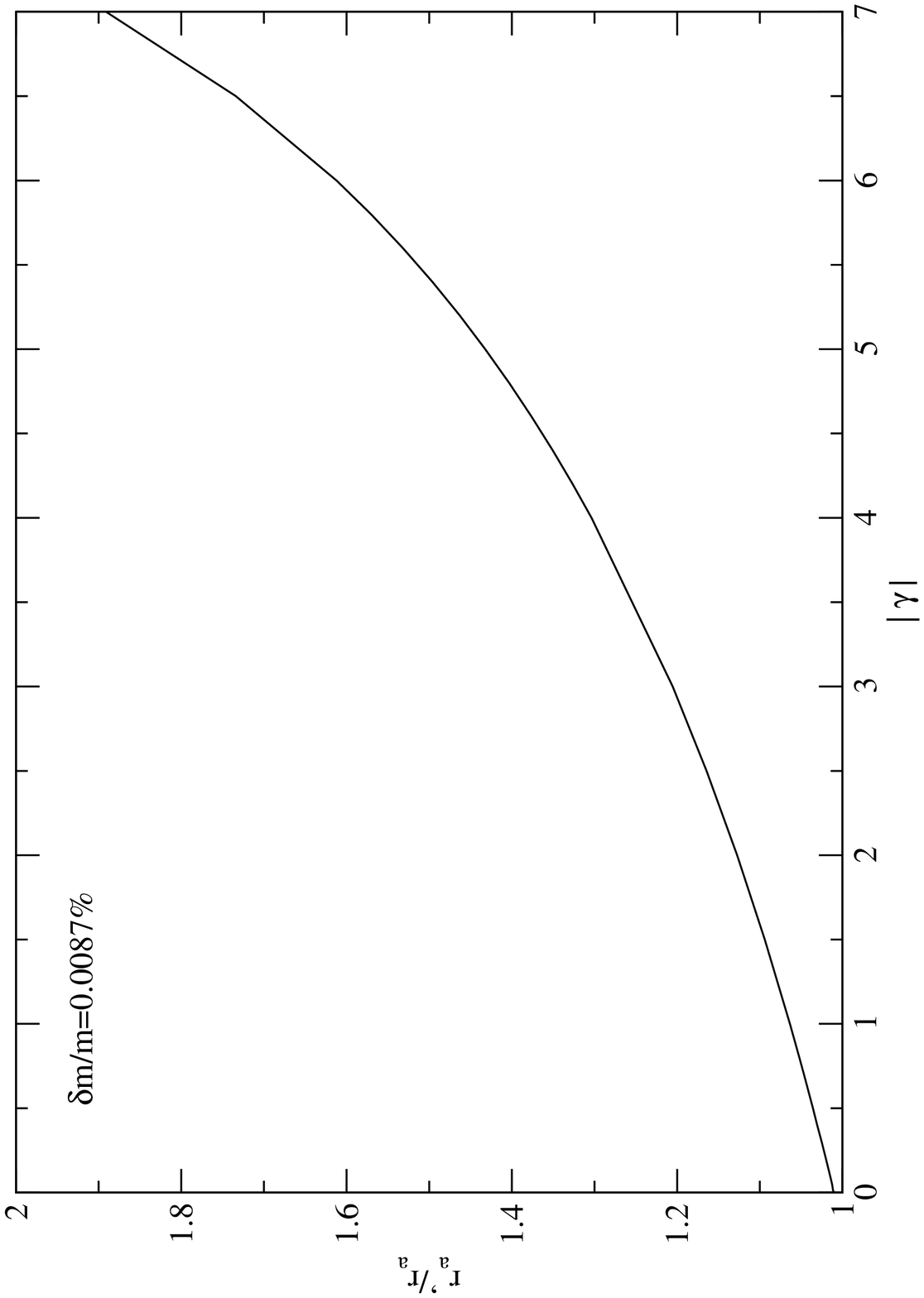}
 \end{center}
 \caption{Ratio of the aphelion increasing as a function of the parameter $\gamma$.}
\end{figure}
\section{\bf Conclusions and comments}
\noindent
The Lagrangian, Hamiltonian and a constant of motion of the gravitational attraction of two bodies were given
when one of the bodies has variable mass and  the dissipative-antidissipative effect of the solar wind is considered. By choosing  the reference system in the
massive body, the system of equations is  reduce  to 1-D problem. Then, the constant of
motion, Lagrangian and Hamiltonian were obtained consistently. A model for comet-mass-variation was given, and with this model, a study was made of  the variation of the period of one cycle of oscillation of the comet when there are mass variation and  dissipation-antidissipation. When mass variation is only considered, the comet trajectory is moving away from the sun,  the mass lost is reduced as the comet is farther away (according to our model), and the period of oscillations becomes bigger. When dissipation-antidissipation is added, this former effect becomes higher as the parameter $\gamma$ becomes higher. \\\\
\noindent
It is important to mention that if instead of loosing mass the body would had winning mass, the period of oscillation of the system would decrease. One can imagine, for example,  a binary stars system where one of the star is winning mass from the interstellar space or from the other star companion. So, due to this winning mass, the period of the star would decrease depending on how much mass the star is absorbing.   \\\\
\newpage
\section{\bf Appendix}
\noindent
Expression for $W_1$ and $W_2$:\\
\begin{eqnarray*}
& &W_1=\frac{Gm_{2-}^2}{m_{2+}^o}\Biggl\{
-\frac{p(p-1)e^{(-4+p)\alpha r}}{2r}+\alpha p E_i(\alpha pr)-2\alpha p(p-1)E_i\bigl((-4+p)\alpha r\bigr)\\
& &\quad+\frac{\alpha p^2(p-1)}{2}E_i\bigl((-4+p)\alpha r\bigr)+
\frac{p(p-1)}{r}\left[e^{(p-3)\alpha r}+3\alpha(1-p)r E_i\bigl((p-3)\alpha r\bigr)\right]\\
& &\quad+\frac{p(p+3)}{2}\left[-\frac{e^{(p-2)\alpha r}}{r}+\alpha(p-2)E_i\bigl((p-2)\alpha r\bigr)\right]\\
& &\quad+\frac{p+2}{r}\left[e^{(p-2)\alpha r}+\alpha(p-1)rE_i\bigl((p-1)\alpha r\bigr)\right]\Biggr\}\\
& &\quad+\frac{l_{\theta}^2}{2m_{2+}^2r^2}\Biggl\{
\frac{p(p-1)}{2}e^{(p-2)\alpha r}-pe^{(p-1)\alpha r}-\alpha p(p-1)e^{(p-2)\alpha r}+\frac{\alpha p(p-1)}{2} e^{p\alpha r}\\
& & \quad+\frac{\alpha^2p(p-1)r}{2}e^{(p-2)\alpha r}+p\alpha r e^{(p-1)\alpha r}-p^2\alpha r e^{(p-1)\alpha r}
-p^2\alpha^2 r^2 E_i\bigl(p\alpha r\bigr)\\
& &\quad-\frac{\alpha^2(p-2)^2p(p-1)r^2}{2}E_i\bigl((p-2)\alpha r\bigr)+p\alpha^2r^2 E_i\bigl((p-1)\alpha r\bigr)\\
& & \quad-2\alpha^2p^2r^2E_i\bigl((p-1)\alpha r\bigr)+p^3\alpha^2 r^2E_i\bigl((p-1)\alpha r\bigr)\Biggr\}
\end{eqnarray*}
$$\eqno(A1)$$
where $m_a$ is the mass of the body at the aphelion, and  we have made the definitions
$$p=\frac{2\gamma}{\alpha m_a}\eqno(A2)$$
and the function $E_i$ is the exponential integral,
$$E_i(z)=\int_{-z}^{\infty}\frac{e^{-t}}{t}dt\eqno(A3)$$
\begin{eqnarray*}
& &W_2=\frac{Gm_{2-}^2}{m_{2+}^o}\Biggl\{
\frac{e^{(q-2)\alpha r}}{r}\left[1+\frac{q(q-1)}{2(m_p+\alpha q)}e^{2q\alpha r}+\frac{2q}{m_p+\alpha q}e^{q\alpha r}\right]\\
& &\quad+q\alpha E_i\bigl(q\alpha r\bigr)
-\frac{q(q-1)e^{2q\alpha r}}{(m_p+\alpha q)^2r}\left[e^{(q-3)\alpha r}-\alpha(q-3)r E_i\bigl((q-3)\alpha r\bigr)\right]\\
& &\quad +\frac{qe^{q\alpha r}}{(m_p+\alpha q) r}\left[e^{(q-3)\alpha r}-\alpha(q-3)r E_i\bigl((q-3)\alpha r\bigr)\right]
-2\alpha E_i\bigl((q-2)\alpha r\bigr)\\
& &\quad+\alpha q E_i\bigl((q-2)\alpha r\bigr)-\frac{q(q-1)\alpha e^{2q\alpha r}}{(m_p+\alpha q)^2}E_i\bigl((q-2)\alpha r\bigr)\\
& &\quad+\frac{q^2(q-1)\alpha e^{2q\alpha r}}{2(m_p+\alpha q)^2}E_i\bigl((q-2)\alpha r\bigr)-
\frac{4\alpha e^{q\alpha r}}{m_p+\alpha q}E_i\bigl((q-2)\alpha r\bigr)\\
& &\quad+\frac{2q^2\alpha e^{q\alpha r}}{(m_p+\alpha q)r}E_i\bigl((q-2)\alpha r\bigr)+
\frac{2}{r}\left[e^{(q-1)\alpha r}-(q-1)\alpha r E_i\bigl((q-1)\alpha r\bigr)\right]\\
& &\quad+\frac{qe^{q\alpha r}}{(m_p+\alpha q)r}\left[e^{(q-1)\alpha r}-(q-1)\alpha r E_i\bigl((q-1)\alpha r\bigr)\right]\Biggr\}\\
& &\quad+\frac{l_{\theta}^2}{2m_{2+}^2(m_p+\alpha q)^q}\Biggl\{
-\frac{q\alpha e^{q\alpha r}}{r}+q^2\alpha^2 E_i\bigl(q\alpha r\bigr)\\
& &+\frac{q(q-1)e^{(3q-2)\alpha r}}{2(m_p+\alpha q)^2r^2}\left[-1+2\alpha r-q\alpha r+(2-q)^2\alpha^2r^2 e^{(2-q)\alpha r} 
E_i\bigl((q-2)\alpha r\bigr)\right]\\
& & -\frac{qe^{(2q-1)\alpha r}}{(m_p+\alpha q)r^2}
\left[-1+\alpha r+q\alpha r+(q-1)^2\alpha^2r^2 e^{(1-q)\alpha r} E_i\bigl((q-1)\alpha r\bigr)\right]\Biggr\}
\end{eqnarray*}
$$\eqno(A4)$$
where $m_p$ is the mass of the body at the perihelion, and we have made the definition
$$q=\frac{2\gamma}{\alpha(m_p-b)}\eqno(A5)$$

\newpage
\section{Bibliography}
\noindent
{\bf 0.} G. L\'opez, L.A. Barrera, Y. Garibo, H. Hern\'andez, J.C. Salazar,\\
\quad and C.A.  Vargas, {\it One-Dimensional Systems and Problems Associated 
\quad with Getting Their Hamiltonians},
\quad Int. Jour. Theo. Phys.,{\bf 43},10 (2004),1.\\ \\
\noindent
{\bf 1.} H. Gylden, 
{\it Die Bahnbewegungen in einem Systeme von zwei K\"orpern in dem Falle, dass die Massen Ver\"anderungen unterworfen sind},\\
\quad Astron. Nachr., {\bf 109}, no. 2593 (1884),1.\\
\quad I.V. Meshcherskii, 
{\it Ein Specialfall des GyldŽn'schen Problems (A. N. 2593)},\\
\quad Astron. Nachr., {\bf 132}, no. 3153 (1893),93.\\
\quad I.V. Meshcherskii, 
{\it Ueber die Integration der Bewegungsgleichungen im Probleme zweier Kšrper von verŠnderlicher Masse},\\
\quad Astron. Nachr., {\bf 159}, no. 3807 (1902),229.\\
\quad E.O. Lovett, 
{\it Note on GyldŽn's equations of the problem of two bodies with masses varying with the time},\\
\quad Astron. Nachr.,{\bf 158}, no. 3790 (1902), 337.\\
\quad J.H. Jeans,
{\it Cosmogonic problems associated with a secular decrease of mass},\\
\quad MNRAS, {\bf 85}, no. 1 (1924),2.\\
\quad L.M. Berkovich, 
{\it Gylden-Mescerskii problem}, \\
\quad Celestial Mechanics, {\bf 24} (1981),407.\\
\quad A.A. Bekov, 
{\it Integrable Cases and Trajectories in the Gylden-Meshcherskii Problem},\\
\quad Astron. Zh., {\bf 66} (1989),135.\\
\quad C. Prieto and J.A. Docobo, 
{\it Analytic solution of the two-body problem with slowly decreasing mass},
\quad Astron. Astrophys., {\bf 318} (1997),657.\\ \\
{\bf 2.} A. Sommerfeld,{\it Lectures on Theoretical Physics}, Vol. I, \\
\quad Academic Press (1964).\\Ê\\
{\bf 3.} A.G. Zagorodny, P.P.J.M. Schram, and S.A. Trigger, 
{\it Stationary Velocity and Charge Distributions of Grains in Dusty Plasmas},\\
\quad Phys. Rev. Lett., {\bf 84} (2000),3594.\\ \\
{\bf 4.} O.T. Serimaa, J. Javanainen, and S. Varr\'o, 
{\it Gauge-independent Wigner functions: General formulation},\\
\quad Phys. Rev. A, {\bf 33}, (1986), 2913. \\ \\
{\bf 5.} H.A. Bethe, 
{\it Possible Explanation of the Solar-Neutrino Puzzle},\\
\quad Phys. Rev. Lett.,{\bf 56}, (1986),1305.\\
\quad E.D. Commins and P.H. Bucksbaum, {\it Weak Interactions of\\
\quad  Leptons  and Quarks}, Cambridge University Press (1983).\\ \\
{\bf 6.} F.W. Helhl, C. Kiefer and R.J.K. Metzler,{\it Black Holes: Theory \\
\quad  and Observation}, Springer-Verlag (1998).\\ \\
{\bf 7.} P.W. Daly, 
{\it The use of Kepler trajectories to calculate ion fluxes at multi-gigameter distances from Comet},\\
\quad Astron. Astrophys., {\bf 226} (1989) 318.\\ \\
{\bf 8.} H. Goldstein,{\it Classical Mechanics}, Addison-Wesley, M.A., (1950).\\ \\
{\bf 9.} G. L\'opez, {\it Partial Differential Equations of First Order and\\
Their Applications to Physics}, World Scientific, 1999.\\ \\
{\bf 10.} F. John,{\it Partial Differential Equations}, Springer-Verlag \\
\quad New York (1974).\\ \\
{\bf 11.} J.A. Kobussen, 
{\it Some comments on the Lagrangian formalism for systems with general velocity-dependent forces},\\
\quad Acta Phys. Austr. {\bf 51},(1979),193.\\ \\
{\bf 12.} C. Leubner, 
{\it Inequivalent lagrangians from constants of the motion},\\
\quad Phys. Lett. A {\bf 86},(1981), 2.\\ \\
{\bf 13.} G. L\'opez, 
{\it One-Dimensional Autonomous Systems and Dissipative Systems}\ ,\\
\quad Ann. of  Phys., {\bf 251},2 (1996), 372.\\ \\
{\bf 14.} G. L\'opez, 
{\it Constant of Motion, Lagrangian and Hamiltonian of the 
 Gravitational Attraction of Two Bodies with Variable Mass },\\
\quad Int. Jour. Theo. Phys., {\bf 46}, no. 4, (2007), 806.\\ \\
{\bf 15.} G. Cevolani, G. Bortolotti and A. Hajduk, 
{\it Debris from comet Halley, comet's mass loss and age}\ , \\
\quad IL Nuo. Cim. C, {\bf 10}, no.5 (1987) 587.\\ \\
{\bf 16.}  J.L. Brandy, 
{\it Halley's Comet: AD 1986 to 2647 BC},\\
\quad J. Brit. astron. Assoc., {\bf 92}, no. 5 (1982) 209.\\ \\
{\bf 17.} D.C. Jewitt, 
{\it From Kuiper Belt Object to Cometary Nucleus: The Missing Ultrared Matter},\\
\quad Astron. Jour.,  {\bf 123}, (2002) 1039. \\Ê\\
{\bf 18.} B.V. Chirikov and V.V. Vecheslavov,
{\it Chaotic dynamics of Comet Halley}\,\\
\quad Astron. Astrophys., {\bf 221},  (1989) 146.\\

\end{document}